\documentclass[aps,prl,twocolumn,groupedaddress,showpacs,floatfix]{revtex4}
\def\bc{\begin{center}}
\def\ec{\end{center}}
\def\be{\begin{equation}}
\def\ee{\end{equation}}

\usepackage{longtable}
\usepackage{epsfig}
\usepackage{psfig}

\begin{document}
\title{Flavor altering excitations of composite fermions}
% Force line breaks with \\

\author{Michael R. Peterson and Jainendra K. Jain}

\affiliation{Department of Physics, 104 Davey Laboratory, The Pennsylvania State
University, University Park, Pennsylvania 16802}

\date{\today}


\begin{abstract}
Past theoretical studies have considered excitations of a given flavor of 
composite fermions across composite-fermion quasi-Landau 
levels.  We show that in general there exists a ladder of flavor 
changing excitations in which composite fermions shed none, some, or all of 
their vortices.  The lowest energy excitations are obtained when the composite 
fermions do not change their flavor, whereas in the highest energy excitations 
they are stripped of all of their vortices, emerging as electrons 
in the final state.  The results are relevant to the intriguing 
experimental discovery\cite{Pinczuk} of Hirjibehedin {\em et al.} 
of coexisting excitation modes of composite fermions of different flavor 
in the filling factor range $1/3>\nu\geq 1/5$. 
\end{abstract}
\maketitle
%}

\maketitle

The quantum fluid of interacting electrons in the lowest Landau level, 
which manifests itself most dramatically through the fractional quantum Hall 
effect\cite{Stormer}, is characterized by the formation of a new class of 
fermions called composite fermions\cite{Jain} (CFs).  
The composite fermion is the bound state of an electron and an even number 
($2p$) of quantized vortices, denoted by $^{2p}$CF.  Composite fermions of 
different flavors (that is, with different numbers of attached vortices)
occur in different filling factor regions, where the filling factor,
$\nu=\rho hc/eB$ is the number of occupied Landau levels (LLs). 
(Here $\rho$ is the two-dimensional density of
electrons and $B$ is the external magnetic field.)
Specifically, $^{2p}$CFs are relevant in the range
$\frac{1}{2p-1}>\nu\geq \frac{1}{2p+1}$.

Light scattering has proved to be a powerful tool for investigating 
the nature of excitations of this quantum fluid 
\cite{RamanPinczuk,RamanKang,RamanDujovne,RamanDavies}.
The present work has been motivated by the 
recent light scattering experiment by Hirjibehedin et al.\cite{Pinczuk},
which investigated the filling factor range $1/3\geq \nu\geq 1/5$,
where the relevant composite fermions carry four vortices
($^4$CFs).  A remarkable aspect of the experimental 
results is that while {\em new} low-energy 
excitations appear for $\nu<1/3$, as expected,  
the {\em old} excitations from $\nu>1/3$ do not disappear but rather 
evolve continuously as the filling factor changes from 
$\nu>1/3$ to $\nu<1/3$; in other words, there is 
a coexistence of excitations of both $^4$CFs and $^2$CFs 
in the low filling factor region.
%They suggest that
%the observation might indicate a coexistence of $^4$CF and
%$^2$CF liquids for $\nu\leq 1/3$.

\begin{figure}
\centerline{\psfig{figure=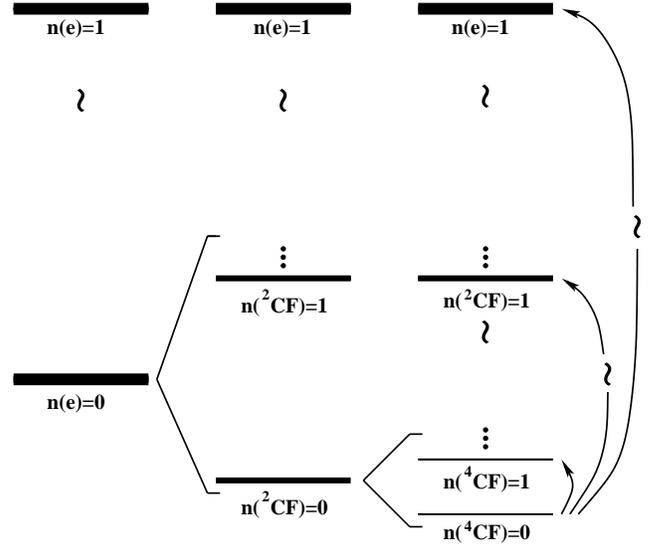,width=3.0in,angle=-90}}
\caption{\label{fig1}
Schematic view of how the lowest electronic Landau level 
splits into quasi-Landau levels of composite fermions carrying 
two vortices ($^2$CFs), and the lowest $^2$CF quasi-Landau level
further splits into $^4$CF quasi-Landau levels.  The (quasi-)Landau level 
index is denoted by n(P) where ``P" is the type of 
relevant particle, namely ``e" (electron), ``$^2$CF," or 
``$^4$CF."  On the rightmost column, the three ladders of 
flavor changing excitations are indicated.  The lowest 
rung in each ladder corresponds to: $\Delta^{4\rightarrow 4}$ from 
n($^4$CF)$=0$ to n($^4$CF)$=1$; $\Delta^{4\rightarrow 2}$ from
n($^4$CF)$=0$ to n($^2$CF)$=1$; $\Delta^{4\rightarrow 0}$ from
n($^4$CF)$=0$ to n(e)$=1$.
}
\end{figure}

We begin by presenting a physical explanation.
A consideration of the filling factor region $\nu>1/3$ 
(Fig.~\ref{fig1}) provides valuable insight 
into this intriguing observation.  Here, we have the lowest electronic LL 
partially occupied, and, as  a result of the repulsive interaction, 
electrons transform into $^2$CFs; the 
lowest electronic LL thus splits into quasi-Landau levels\cite{comment} 
(QLLs) of $^2$CFs (middle column of Fig.~\ref{fig1}).
Now, the {\em intra-}LL excitations of electrons are
described as {\em inter-}QLL excitations of $^2$CFs.
However, the system obviously still supports the inter-{\em electronic}
LL excitation, namely the Kohn mode.
We see in this simple case the coexistence
of excitation modes at two different energy scales:
the low energy $\Delta^{2\rightarrow 2}(k)$ excitations conserve 
the CF flavor, 
whereas the high energy $\Delta^{2\rightarrow 0}(k)$ excitations involve 
transformation of $^2$CF into $^0$CF (electron). 
In the former case, the composite fermions maintain 
their integrity during the transition,
whereas in the latter case they leave their 
vortices behind in the lowest electronic Landau level,
thus converting into excited {\em electrons} in the second level. 

\begin{figure}
\centerline{\psfig{figure=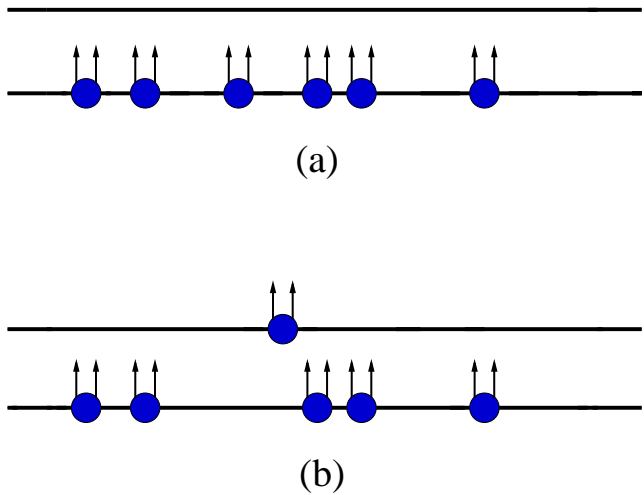,width=3.5in,angle=-90}}
\caption{\label{fig2}
The $^2$CF basis states included in our study to 
investigate the excitations $\Delta^{4\rightarrow 4}$ and
$\Delta^{4\rightarrow 2}$ in the filling factor range 
$1/3> \nu \geq 1/5$.  (The $^2$CFs are shown as electrons carrying two 
arrows, and the horizontal lines depict their quasi-Landau levels.) 
These are constructed from the analogous basis states for electrons.  
Keeping only states of the type given in (a)  
produces the $\Delta^{4\rightarrow 4}$ excitations shown by 
empty circles in Fig~\protect\ref{fig3}.
}
\end{figure}

For $\nu<1/3$ the situation is 
analogous.  The lowest $^2$CF QLL undergoes 
a further splitting into QLLs of $^4$CFs, as shown in 
the rightmost column of Fig.~\ref{fig1}.  Although we have only $^4$CFs 
in the ground state, there are three ladders of excitation: 
(i) $\Delta^{4\rightarrow 4}(k)$ excitations, which 
were not present for $\nu>1/3$;
(ii) $\Delta^{4\rightarrow 2}(k)$ excitations involving a change of 
$^4$CFs into $^2$CFs, which are a continuous evolution of  
$\Delta^{2\rightarrow 2}(k)$ excitations for $\nu>1/3$;
and (iii) $\Delta^{4\rightarrow 0}(k)$, which are similarly 
close cousins of $\Delta^{2\rightarrow 0}(k)$ and 
$\Delta^{0\rightarrow 0}(k)$.  This is our physical explanation for 
the experimental result of Ref.~\onlinecite{Pinczuk} showing
a coexistence of several kinds of excitations.

\begin{figure}
\centerline{\psfig{figure=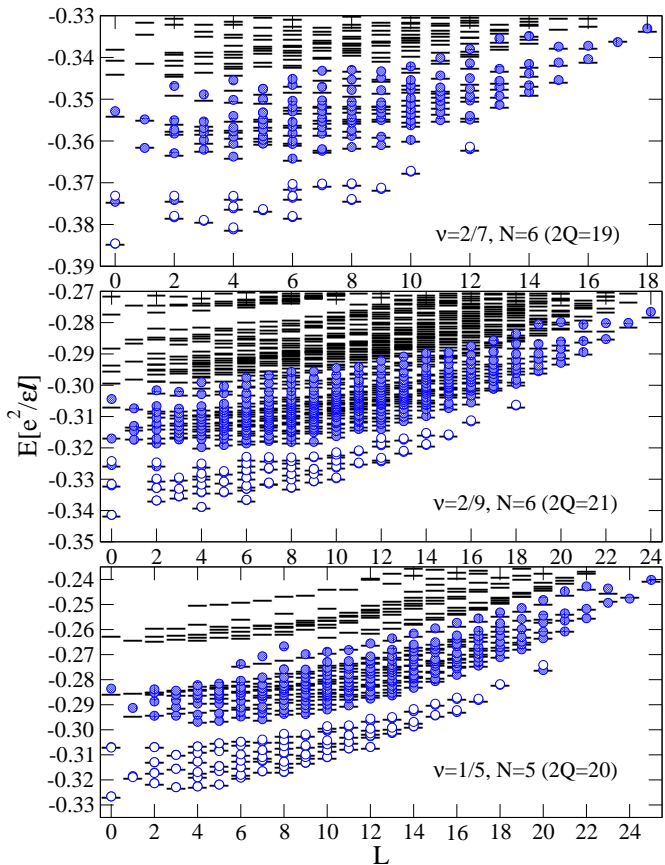,width=3.5in,angle=0}}
\caption{\label{fig3}
Numerical results are shown for filling factors $\nu=2/7$, 2/9, and 
1/5 for $N=6$, 6, and 5 particles, respectively.  
The spherical geometry is used in the calculation, with the magnetic 
flux through the surface equal to $2Qhc/e$.
The filled circles show the spectrum calculated from 
the CF theory, obtained, as explained in the text, by diagonalization 
in the $^2$CF basis shown schematically in Fig.~\protect\ref{fig2}.
The empty circles are the CF states obtained when only the basis 
states of the type given in Fig.~\protect\ref{fig2}a are kept.
Note that there is a filled circle underneath each empty circle
(sometimes partly visible).  Dashes show the exact spectrum.   
$L$ is the total orbital angular momentum,
and the energy is reported in units of $e^2/\epsilon l$ where 
$\epsilon$ is the dielectric constant 
of the host semiconductor and $l=\sqrt{\hbar c/eB}$ is 
the magnetic length.  The Monte 
Carlo uncertainty is smaller than the symbol size, hence not shown 
explicitly.  The energies have been corrected for the finite 
size deviation in the electron density through a multiplicative 
factor $\sqrt{\rho/\rho_N}=\sqrt{2Q\nu/N}$.
}
\end{figure}

To test these ideas quantitatively through a microscopic calculation, 
we need to construct a model that can deal with 
both $\Delta^{4\rightarrow 4}$ and
$\Delta^{4\rightarrow 2}$ excitations simultaneously. 
The key observation that makes this possible 
is that the {\em intra-}QLL excitations of a given flavor 
of composite fermions are the {\em inter-}QLL excitations of higher
order composite fermions.  For example, at filling factors 
of the type $\nu=n/(2n+1)$ (with $n$ integer),  which are 
described in terms of $^2$CFs, the intra-LL excitations of 
electrons ($^0$CFs) are inter-QLL excitations of $^2$CFs.  
If we diagonalize this problem within the lowest electronic 
LL, we will miss the Kohn mode.  In order to keep 
this mode, we must expand our basis space to allow at least one 
electron to occupy the second electronic LL; then we get not only
all the $\Delta^{2\rightarrow 2}$ excitations, but also the 
lowest energy $\Delta^{2\rightarrow 0}$ excitations.  This idea can be 
transported straightforwardly to the filling factor region $1/3>\nu\geq 1/5$
by the vortex attachment trick of the CF theory, which relates 
$\nu^*$ to $\nu=\nu^*/(2\nu^*+1)$.
We first construct an electron basis at $\nu^*$, wherein we keep all 
basis states in which either all electrons are in the lowest LL 
or one electron occupies the second LL.  By attaching two vortices 
to each particle, we generate from it, following methods outlined below,
a $^2$CF basis at $\nu$ 
shown schematically in Figs.~\ref{fig2}a and ~\ref{fig2}b.  
We then diagonalize the interaction Hamiltonian 
in this basis.  An explicit calculation for $N=5$ and 6 for 
certain filling factors in the range $1/3>\nu\geq 1/5$ 
gives the spectrum in Fig.~\ref{fig3} shown by filled circles.

In order to identify the states at $\nu$ that correspond to the Kohn mode
at $\nu^*$, we repeat the above calculation without the 
Kohn mode (that is, by working strictly within the lowest 
LL at $\nu^*$; Fig.~\ref{fig2}a).  We find that this gives a subset 
of the previous states,
shown by empty circles in Fig.~\ref{fig3}.  (All empty 
circles sit on filled circles; for a given 
eigenstate, the energy of the filled circle is slightly lower than that
of the empty circle because of the greater variational freedom in the 
expanded basis, but the difference is so small in most cases that the 
empty circles fully cover the filled ones.)  The states not covered by 
empty circles are therefore {\em inter-}QLL excitations of $^2$CFs, 
that is, $\Delta^{4\rightarrow 2}$ excitations.  

In short, our approach is to model the $1/3>\nu\geq 1/5$ region 
in terms of $^2$CFs -- as opposed to
$^4$CFs, which would be the natural choice if we were only interested in
the lowest energy excitations.  Then, the {\em intra}-QLL excitations of
$^2$CFs give us the {\em inter}-QLL excitations of
$^4$CFs, namely $\Delta^{4\rightarrow 4}$; at the same
time the {\em inter}-QLL excitations of $^2$CFs 
correspond to the $\Delta^{4\rightarrow 2}$ excitations.

There is a remarkable separation of energy scales 
in Fig.~\ref{fig3} between the inter- and intra-QLL 
excitations of $^2$CFs, which is entirely consistent with the physics 
described above.  A similar effect exists  
for the inter- and intra-LL excitations of {\em electrons}, 
which are governed by different 
energies, namely the cyclotron energy and the interaction energy.
For $^2$CFs, however, both the inter- and intra-QLL energies 
are determined by the interaction energy, and the emergence of two 
energy scales is not obvious {\em a priori}. 
Fig.~\ref{fig3} also shows exact diagonalization results (dashes).  
There is an excellent agreement between the dashes and the circles,
further confirming the validity of our picture.  
In particular, the exact spectra also show two distinct energy scales.
(In fact, once we understand the physics, we can identify the $4\rightarrow 4$ 
and $4\rightarrow 2$  energies directly from the {\em exact} spectrum, 
which is what we do below.)
Still higher energy states can be obtained within the framework of the 
CF theory by exciting more than one composite fermion
to higher CF quasi-Landau levels, but our interest here is in the lowest 
excitation at each flavor changing ladder.

We next compare our results with experiment.
In an ideal situation, inelastic light scattering only probes
the excitations close to zero wave vector, but slight disorder,
which breaks wave vector conservation, has been found to expose 
energies at critical points (extrema) in the CF exciton dispersion
\cite{RamanPinczuk,RamanKang,RamanDujovne}.
In particular, the zero wave vector mode, the minimum energy 
neutral excitation (roton),  and the 
excitation that contains a far separated particle hole pair of composite 
fermions (which gives the transport gap measured 
in experiments) have been observed in Raman experiments.  We will concentrate
on the first two, which will be labeled $\Delta(0)$ and $\Delta(R)$,
respectively.  The transport gap is roughly the same as $\Delta(R)$
for our small systems.

\begin{figure}
\centerline{\psfig{figure=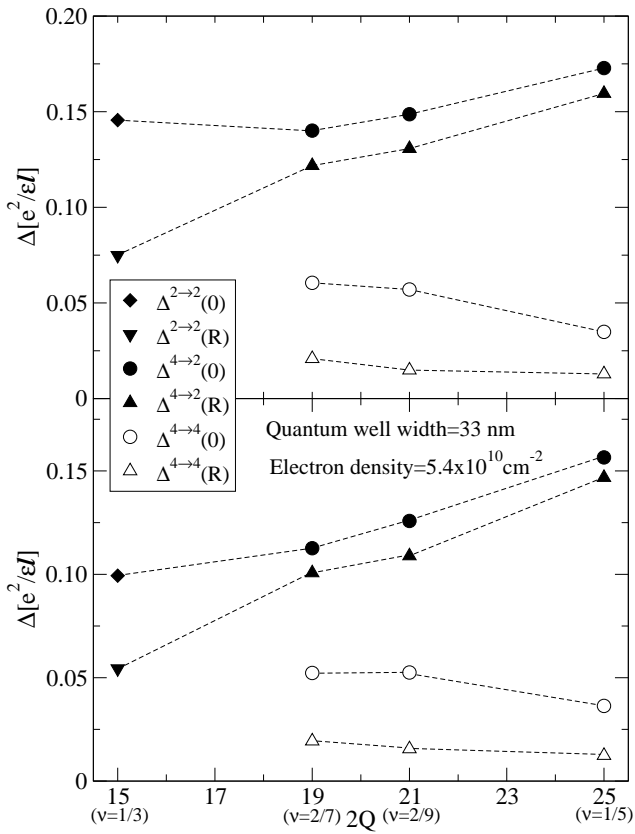,width=3.5in,angle=0}}
\caption{\label{fig4}
Excitation energies $\Delta^{2p\rightarrow 2p'}$ of the 
zero wave vector and roton modes are shown as a function of $2Q$ (or 
magnetic field).  The filling factor corresponding to 
a particular magnetic field strength is given in parenthesis along 
the horizontal axis.  The upper panel gives the 
energies for zero thickness, whereas the lower 
panel incorporates corrections due to finite thickness using 
parameters appropriate for the experimental sample 
of Ref.~\protect\onlinecite{Pinczuk} (quantum well width = 33 nm and density 
$= 5.4 \times 10^{11}$ cm$^{-2}$).  All results are for $N=6$ particles, 
and the dashed lines are a guide to the eye.
}
\end{figure}

The lowest energy intra- and inter-$^2$CF QLL excitations 
can be identified straightforwardly from the spectra of Fig.~\ref{fig3}.
The upper panel of Fig.~\ref{fig4} shows the energies 
$\Delta^{4\rightarrow 4}(0)$,
$\Delta^{4\rightarrow 4}(R)$, and their $4\rightarrow 2$ 
counterparts.  While the $4\rightarrow 2$ 
modes evolve continuously out of the $2\rightarrow 2$ modes 
at $\nu\geq 1/3$, the $4\rightarrow 4$ modes are new.  The qualitative 
behavior is in agreement with experiment.  There 
are, however, several difficulties in making a quantitative comparison 
with experiment.

First, the calculations are performed for 
a strictly two dimensional system, whereas the experimental 
sample is a square quantum well of width 33 nm.  
We incorporate finite thickness corrections into our 
calculation through a self consistent local density 
approximation\cite{Ortalano}.  The lower panel of 
Fig.~\ref{fig4} depicts the various energies after including 
modification in the effective two-dimensional interaction 
due to finite extent of the electron wave function in the 
transverse direction.

The other problems are more severe.  
Strictly speaking, one must calculate thermodynamic limits 
before comparing theoretical calculations to experiment.  
In the past, energy gaps have been found to vary substantially
as a function of $N$.  (For example, the transport
gap at $\nu=2/5$ is reduced from
0.13 for $N=4$ to  $0.058$ in the thermodynamic limit
\cite{Manual}.) However, because of the rather large dimension of 
the CF basis, we are not able to go to systems bigger 
than $N=6$ in a numerically stable manner.
The calculation also neglects disorder, which 
is often a source of significant quantitative correction.

Next we compare our theoretical results for the roton energy  
with those in the experiment of Ref.~\onlinecite{Pinczuk} for the  
filling factor range $1/3\geq \nu \geq 1/5$.  For $\Delta^{4\rightarrow 4}(R)$ 
the calculated values are approximately $\sim0.2$ meV in this 
range (taking parameters of the experiment in Ref.~\onlinecite{Pinczuk}), 
decreasing slightly with 
increasing magnetic field.  The experimental values start at 
$\sim0.2$ meV and decrease with magnetic field to  
$\sim0.11$ meV at $\nu=1/5$.  
For $\Delta^{4\rightarrow 2}(R)$ mode the theoretical  
value starts at approximately $\sim0.6$ meV and increases with magnetic field 
to $\sim2.0$ meV.  The experimental value starts at $\sim0.45$ meV and 
goes on increasing with magnetic field to $\sim0.7$ meV.  
The theory is in qualitative agreement with experiment, both 
in regard to (i) the continuity of some modes across $\nu=1/3$ 
and appearance of new modes for $\nu<1/3$, and  (ii) the observed 
trends as a function of the filling factor.  In light of the rather 
small size of the numerical system,
we find the level of quantitative agreement to be satisfactory. 

Finally, we briefly discuss the calculational method.  
It involves many steps, which will only be outlined here for lack 
of space; more details can be found in the 
literature\cite{Manual,Mandal,PetersonJain,MPJ}.  
We have used the standard 
spherical geometry\cite{Haldane,WuYang}  which considers $N$ electrons 
on the surface of a sphere with a magnetic monopole 
of strength $Q$ placed at the center producing a radial magnetic field.  
The magnetic flux through the sphere is $2Q\phi_0$ ($\phi_0=hc/e$ 
is called the magnetic flux quantum); $2Q$ is
an integer due to Dirac's quantization condition.
The single particle eigenstates are the 
monopole harmonics\cite{WuYang} $Y_{Qnm}(\Omega_j)$ 
where $n=0,1,\ldots$ is the LL 
index, $m=-(Q+n),-(Q+n)+1,\ldots,(Q+n)$ labels 
the $2(Q+n)+1$ degenerate states in the $n$th LL, 
and $\Omega_j$ represents the location on the sphere of 
electron $j$ with the usual coordinates.  Since the degeneracy 
in successive LLs is not equal in this geometry, the ratio $N/2Q$ is not 
exactly equal to the filling factor for finite systems;
the filling factor is defined as $\nu=\lim_{N\rightarrow\infty}N/2Q$.

According to the CF theory, $N$ strongly interacting electrons 
at monopole strength $Q$ are transformed into $N$ weakly 
interacting composite fermions at monopole strength $Q^*=Q-p(N-1)$.  
For a given $Q$ in the range $1/3 > \nu \geq 1/5$, we first 
consider $Q^*=Q-N+1$, which lies in the range $1 > \nu \geq 1/3$.
We diagonalize the electron problem at $Q^*$ (allowing at most 
one electron in the second LL) to obtain eigenstates with 
definite total orbital angular momentum $L$.  
Denoting this basis by $\{\Phi_{Q^*}\}$, 
we construct a basis at $Q$ given by $\{P_{LLL}\Phi_1^2\Phi_{Q^*}\}$,
where $\Phi_1$ is the wave function of one filled Landau level.
$P_{LLL}$ denotes projection into the lowest electronic 
LL, which can be accomplished using the 
methods discussed previously\cite{Manual}.
The basis at $Q$ obtained in this manner is not orthogonal.
Following the techniques in Ref.~\onlinecite{Mandal}, we 
use the Gram-Schmidt procedure to obtain an orthogonal 
basis, and then diagonalize the interaction Hamiltonian 
in this basis.  The various projections and Hamiltonian 
matrix elements require evaluation of multi-dimensional 
integrals, which is accomplished numerically by Monte 
Carlo.  The uncertainty is determined by 
the standard deviation in the eigenvalues over different 
Monte Carlo runs.

A technical point in regard to our treatment of finite width 
should be noted.  Following Ref.~\onlinecite{Xie}, we 
determine the interaction pseudopotentials\cite{Haldane} 
in the infinite planar geometry, and feed them  
into exact calculations in the spherical geometry.  
For an infinite system, this of course gives the same 
result as doing the calculation entirely within the 
spherical geometry, but for finite systems the two 
methods are different.
For zero thickness, the use of planar pseudopotentials, 
as opposed to the spherical pseudopotentials, 
makes only a slight deviation (within 6\% for our calculations)
in the excitation energies; we expect that the method works equally 
well for finite width systems as well.

We have only considered in this 
work filling factors where the ground state is incompressible.  
It would be of interest to investigate the filling factor dependence 
of the various excitations energies.

This work was supported in part by the National Science Foundation under grants
no. DGE-9987589 (IGERT) and DMR-0240458.  We are grateful to the High Performance
Computing (HPC) Group led by V. Agarwala, J. Holmes, and J. Nucciarone, at the
Penn State University ASET (Academic Services and Emerging Technologies) for
assistance and computing time with the LION-XL cluster.


\begin{references}
\bibitem{comment}The quasi-Landau-like levels of composite fermions must 
not be confused with the ordinary Landau levels of electrons.
\end{references}
\end{document}